\documentstyle[12pt,epsf]{article}
\advance\oddsidemargin by -1in
\advance\evensidemargin by -1in
\advance\topmargin  by -0.5in
\renewcommand{\pmod}[1]{\ (\bmod\ #1)}
\newcommand{\qb}{{\cal B}}
\newcommand{\BQP}{{\mathrm BQP}}
\newcommand{\calA}{{\cal A}}
\newcommand{\calH}{{\cal H}}
\newcommand{\calF}{{\cal F}}
\newcommand{\calG}{{\cal G}}
\newcommand{\calL}{{\cal L}}
\newcommand{\calM}{{\cal M}}
\newcommand{\LL}{{\bf L}}
\newcommand{\RR}{{\bf R}}
\newcommand{\CC}{{\bf C}}
\newcommand{\Image}{\mathop{\rm Im}\nolimits}
\newcommand{\binary}[1]{\overline{\strut#1}}
\newcommand{\bra}[1]{\langle #1|}
\newcommand{\ket}[1]{|#1\rangle}

\newcommand{\occup}{n}
\newcommand{\qswap}{\leftrightarrow}
\newcommand{\fswap}{\Leftrightarrow}
\newcommand{\qapp}[1]{[#1]}
\newcommand{\fapp}[1]{\{#1\}}
\newcommand{\mapp}[1]{\{\!\{#1\}\!\}}
\newcommand{\qext}[1]{[{\leftarrow}#1]}
\newcommand{\fext}[1]{\{\!{\Leftarrow}#1\!\}}
\newcommand{\ppext}[1]{\widetilde{#1}}
\newcommand{\locless}[1]{\underrel{#1}{<}}
\newcommand{\underrel}[2]{\mathrel{\mathop{#2}\limits_{#1}}}

\title{Fermionic quantum computation}
\author{Sergey~B.~Bravyi\\
{\it L.~D.~Landau Institute for Theoretical Physics,}\\
{\it Kosygina St.~2, Moscow, 117940, Russia,}\\
{\it serg@itp.ac.ru}\\
and\\
Alexei~Yu.~Kitaev\thanks{On leave from  L.~D.~Landau Institute
for Theoretical Physics}\\
{\it Microsoft Research}\\
{\it Microsoft, \#113/2032, One Microsoft Way,}\\
{\it Redmond, WA 98052, U.S.A.}\\
{\it kitaev@microsoft.com}
}
\date{March 31, 2000} 
 
\begin{document}
\maketitle

\begin{abstract}
We define a model of quantum computation with local fermionic modes (LFMs) ---
sites which can be either empty or occupied by a fermion. With the standard
correspondence between the Foch space of $m$ LFMs and the Hilbert space of $m$
qubits, simulation of one fermionic gate takes $O(m)$ qubit gates and vice
versa. We show that using different encodings, the simulation cost can be
reduced to $O(\log m)$ and a constant, respectively. Nearest-neighbors
fermionic gates on a graph of bounded degree can be simulated at a constant
cost. A universal set of fermionic gates is found. We also study computation
with Majorana fermions which are basically halves of LFMs. Some connection to
qubit quantum codes is made.
\end{abstract}
 
\section*{Introduction}
The notion of {\em locality}\/ plays the key role in the definition of
computation process. The same basic principles apply to classical computers
and the circuit model of quantum computation~\cite{Feynman,Deutch}:
\begin{enumerate}
\item The computer consists of small pieces, or cells (bits, qubits,
qutrits or something else). 
\item It is allowed to operate on few cells at a time.
\item All cells are identical, so each operation has a model which can be
applied to different sets of cells. We call such a model a {\em gate} (like
the CNOT gate) while the operation itself is called a {\em gate application}
(like CNOT applied to qubits $5$ and $8$).
\end{enumerate}
In fact, the main difference between classical computation and quantum
computation is the concrete interpretation of these postulates. The standard
quantum interpretation is as follows.
\begin{itemize}
\item[1i.] Each cell is described by a Hilbert space of small dimensionality.
(Without loss of generality, these spaces are two-dimensional, in which case
the cells are called qubits). The Hilbert space of the entire computer is the
tensor product of the spaces associated to the individual cells.
\item[2i.] Each operation is described by a unitary operator which is the
tensor product of some operator $U$, acting on the selected qubits, and the
identity operator acting on the rest of the system.
\item[3i.] A $p$-qubit gate can be defined as $U$ acting on some standard
$2^p$-dimensional space (which is the same as a standard set of $p$ qubits).
\end{itemize}

This interpretation might be perfect from the complexity-theoretic point of
view, but it does not necessarily correspond to physics. At the fundamental
level, Fermi systems do not satisfy the condition~(2i). Hence using fermions
as carriers of quantum information~\cite{Lloyd} should be considered as a
different computation model, although it is equivalent to the standard one in
a certain sense. At the macroscopic level, collective quantum degrees of
freedom (or excitations, such as anyons~\cite{Pfaffian,Kit1}) do not even
satisfy the condition~(1i). In all such cases, it makes sense to abstract the
nontrivial locality properties from physical details. This will lead us to a
definition of ``fermionic gates'', ``anyonic gates'' and other quantum
computation models which deserve careful study.

What can one expect from alternative models of quantum computation? It is very
unlikely that any physical system would provide more computational power than
the standard quantum computation model has. (This might be wrong for quantum
gravity but here we can only guess). So, the alternative models should be
polynomially equivalent to the standard one. There are indeed several results
which support this statement. Firstly, D.\,Abrams and S.\,Lloyd~\cite{AL} have
shown that a system of $m$ local fermionic modes (i.e.\ sites which can be
empty or occupied) can be simulated on a quantum computer in such a way that
one fermionic gate takes $O(m)$ qubit operations. In some cases (one of the
assumptions is that the number of particles is conserved) faster simulation is
possible. We will extend this result by showing that in the general case (when
the number of fermions is conserved only modulo $2$) each fermionic gate can
be simulated by $O(\log m)$ qubit gates. Secondly, TQFT computation (which is
more general than quantum computation with anyons) can be simulated in
polynomial time on an ordinary quantum computer~\cite{TQFT}.

Thus alternative quantum computation models do not generate new computational
classes. Rather, they provide new descriptions of the standard class $\BQP$
(the class of problems that are solvable on a quantum computer in polynomial
time). These descriptions may be useful to find new quantum algorithms,
error-correcting codes, fault-tolerant procedures, or to prove that $\BQP$ is
contained in some other computational classes. They may also open the door to
new physical implementations of a quantum computer.

\section{A more general notion of locality}\label{sec_locality}

Why isn't the standard interpretation of locality good in all cases? The
answer is in the way we describe quantum evolution. The Hilbert space and
state vectors are very convenient tools but they are not directly related to
physically observable things. Operators are much more ``real'' as they allow
one to describe interaction between the system and the rest of the world. In
fact, the causality principle is usually stated in terms of operators:
``operators in spatially separated points commute''. Our definition of
locality will be in the same spirit.

A quantum system can be characterized by a finite-dimensional
$C^*$-algebra\footnote{$C^*$-algebra is a generalization of the algebra of
linear operators on a Hilbert space. The properties of operator addition,
multiplication, Hermitian conjugation and the operator norm are axiomatized in
a certain way. However, instead of using the axioms, we will rely on a
characterization of finite-dimensional $C^*$-algebras given below.}  $\calG$
whose elements are called ``physical operators''. As a matter of fact, they
{\em are} operators on a suitable Hilbert space. Indeed, it is a theorem that
$\calG$ can be represented as $\bigoplus_{j}\LL(\calH_j)$, where $\LL(\calL)$
stands for the algebra of operators on the space $\calL$. Hence $\calG$ acts
on $\calH=\bigoplus_{j}\calH_j$. (In the case of fermions,
$\calH=\calH_0\oplus\calH_1$ is the Foch space split into the subspaces
corresponding to an even and odd number of particles; $\calG$ is the algebra
of operators which preserve the parity).

To define locality, we will assume that the system is associated with some set
of sites $M$. The following properties are postulated:
\begin{enumerate}
\item For each subset of sites $S\subseteq M$, there is a $C^*$-subalgebra
$\calG(S)\subseteq\calG$. Elements of $\calG(S)$ are called {\em physical
operators acting on $S$}. We require that $\calG(M)=\calG$,\,
$\calG(\emptyset)=\CC \cdot I$ (where $\CC$ is the algebra of complex numbers,
$I$ is the unit element of $\calG$), and $\calG(S)\subseteq\calG(S')$ if
$S\subseteq S'$.
\item If $S_1\cap S_2=\emptyset$ then any two operators $X_1\in\calG(S_1)$ and
$X_2\in\calG(S_2)$ commute.
\end{enumerate}

The concept of unitarity is well defined in this setting: an element
$U\in\calG(M)$ is called {\em unitary} if $UU^\dag=U^\dag U=I$ (the
operation $\dag$ is a part of the $C^*$-algebra structure). Note that
nonunitary elements of $\calG(M)$ also have physical meaning because they can
be used to construct a unitary operator on a larger space $\calH\otimes\calL$,
where $\calL$ represents some external system (e.\,g.\ a measurement
device). Such an operator generally has the form $U=\sum_k A_k\otimes B_k$\,
($A_k\in\calG(M)$), i.\,e.\ $U\in\calG(M)\otimes\LL(\calL)$.

Thus we have given a more general interpretation of the locality postulates 1
and 2 which were discussed in the introduction. (We put aside the postulate 3
here).

\section{Local fermionic modes}

Consider $m$ sites (numbered $0$ through $m-1$) each of which can be either
empty or occupied by a spinless fermionic particle. Such sites will be called
{\em local fermionic modes (LFMs)}. The Hilbert space $\calH$ of this system,
known as {\em Foch space}\/, is spanned by $2^m$ basis vectors
$\ket{\occup_{0},\ldots,\occup_{m-1}}$, where $\occup_j=0,1$ is the occupation
number of the $j$-th site.  Everything related to fermions can be expressed in
terms of annihilation and creation operators $a_j,a^\dag_j$,\,
($j=0,\ldots,m\!-\!1$). The operator $a_j$ acts on basis vectors as follows:
\begin{equation}
\begin{array}{l}
\displaystyle
a_j\, \ket{\occup_{0},\ldots,\occup_{j-1},1,\occup_{j+1},\ldots,\occup_{m-1}}
\,=\, \bigl(-1\bigr)^{\sum_{s=0}^{j-1}\occup_s}\,
\ket{\occup_0,\ldots,\occup_{j-1},0,\occup_{j+1},\ldots,\occup_{m-1}} , \\
a_j\, \ket{\occup_{0},\ldots,\occup_{j-1},0,\occup_{j+1},\ldots,\occup_{m-1}}
\,=\, 0;
\end{array}
\end{equation}
$a_j^\dag$ is the Hermitian conjugate. Note that the definition depends on
the order of LFMs! (One may rather say that the basis depends on the order
while $a_j,a^\dag_j$ do not, since all relations between them are
permutation-invariant). The annihilation and creation operators generate the
algebra $\bar{\calF}=\LL(\calH)$ and have the following commutation
rules:
\begin{equation}
\begin{array}{rcl}
a_j a_k + a_k a_j &=& 0,\\
a_j^\dag a_k^\dag + a_k^\dag a_j^\dag &=& 0,\\
a_j a_k^\dag + a_k^\dag a_j &=& \delta_{jk}.
\end{array}
\end{equation}

The Hilbert space of $m$ LFMs splits into two parts:
$\calH=\calH_0\oplus\calH_1$, where ``0'' and ``1'' refers to the total
fermionic parity $\occup=\sum_{j=0}^{m-1}\occup_j\pmod{2}$. {\em Physical
operators} are those which preserve the parity. Note that the Hamiltonian of a
real Fermi system always satisfies this condition,\footnote{In electon
systems, the Hamiltonian also preserves the electric charge, so terms with
different numbers of $a_j$ and $a_j^\dag$ are usually forbidden. Our model is
mostly relevant to superconductors where the total charge {\em of excitations}
is not conserved, so terms like $a_ja_k$ appear in the effective Hamiltonian.}
unlike the operators $a_j,a_j^\dag$ alone.  The {\em algebra of physical
operators}\/ $\calF=\LL(\calH_0)\oplus\LL(\calH_1)$ is spanned by products of
even number of $a_j,a_j^\dag$. (The notation $\calG$ in the previous section
referred to the general case whereas $\calF$ is reserved for LFMs).

Let $S\subseteq\{0,\ldots,m\!-\!1\}$ be a set of LFMs.  {\em Physical
operators on S} are linear combinations of even products of $a_j,a_j^\dag$,\,
$j\in S$. The algebra of such operators is $\calF(S)=\bar{\calF}(S)\cap\calF$,
where $\bar{\calF}(S)\subseteq\bar{\calF}$ is generated by $a_j,a_j^\dag$,\,
$j\in S$. The conditions (1) and (2) from Sec.~\ref{sec_locality} are
obviously satisfied. Moreover, $\calF(S_1)$ commutes with $\bar{\calF}(S_2)$
if $S_1\cap S_2=\emptyset$.

The occupation number $\occup_j$ is an eigenvalue of the operator
$\hat\occup_j=a_j^\dag a_j\in\calF(\{j\})$, which means it can be measured
locally (by acting on the $j$-th LFM and some external device). The occupation
number can not be changed locally, though.

Here are some examples of unitary operators acting on one or two LFMs:
$\exp(i\beta a_j^\dag a_j)$ (action by an external potential), $\exp(i\beta
a_j^\dag a_j a_k^\dag a_k)$ (two-particle's interaction), $\exp(i(\gamma
a_j^\dag a_k +\gamma^* a_k^\dag a_j))$ (tunneling) and $\exp(i(\gamma a_k
a_j+\gamma^* a_j^\dag a_k^\dag))$ (interaction with a superconductor). We will
show that these operators (for all or for some particular values of
$\beta\in\RR$ and $\gamma\in\CC$) form a universal set, i.e.\ any unitary
operator can be represented as a composition of these ones to any given
precision, using ancillas.

In our computation model we allow to use ancillas in the state $\ket{0}$. (To
be more accurate, we should say that the input state is padded by some number
of zeros to the right. Actually, the order does not matter in this case). If
we speak about implementing a unitary operator, the ancillas must return to
the state $\ket{0}$ by the end of the procedure. As is usual, this
restriction does not apply to computing a Boolean function which proceeds as
follows. One starts from a basis vector $\ket{\occup_0,\ldots,\occup_{m-1}}$
representing the input data, adds some ancillas, applies some sequence of
local unitary operators and reads the result by measuring some of the
occupation numbers.

\section{Relation between LFMs and qubits}\label{sec_fvq}

The Hilbert space of $m$ LFMs can be identified with the Hilbert space of $m$
qubits $\qb^{\otimes m}$ (where $\qb$ stands for the two-dimensional space
$\CC^2$ endowed with the standard basis $\bigl\{\ket{0},\ket{1}\bigr\}\,$):
\begin{equation}\label{fvq}
\ket{\occup_0,\occup_1,\ldots,\occup_{m-1}} \equiv
\ket{\occup_0}\otimes\ket{\occup_1} \otimes \cdots \ket{\occup_{m-1}},\qquad\
\occup_j=0,1.
\end{equation}
Measurement of $\occup_j$ is the same as eigenvalue measurement of
$\sigma^z_j$. A physical fermionic operator corresponds to a qubit operator
which preserves the parity, i.e.\ commutes with
$\prod_{j=0}^{m-1}\sigma^z_j$. However, ``applying a gate to a set of LFMs''
is very different from ``applying a gate to a set of qubits''.

Let $X$ be a parity-preserving $p$-qubit operator acting on qubits numbered
$0$ through $p-1$. Applying it to the qubits $j_0,\ldots,j_{p-1}$ is a
straightforward procedure. The Hilbert space of $m$ qubits $\qb^{\otimes m}$
can be identified with $\qb^{\otimes p}\otimes\qb^{\otimes(m-p)}$ by the qubit
permutation $P:\,\ket{\occup_0,\ldots\occup_{m-1}}\mapsto
\ket{\occup_{j_0},\ldots\occup_{j_{p-1}}}\otimes\ket{\mbox{the other}\,
\occup_j}$. Then the action of $X$ is defined as follows:
\[
X\qapp{j_0,\ldots,j_{p-1}}\,=\,P^{-1}(X\otimes I_{\qb^{\otimes(m-p)}})P.
\]

If we want to apply $X$ to the LFMs $j_0,\ldots,j_{p-1}$, the procedure is
different. First, we should expand $X$ into products of
$a_0,\ldots,a_{p-1},a_0^\dag,\ldots,a_{p-1}^\dag$. Then we replace each
$a_r$ by $a_{j_r}$ and each $a_r^\dag$ by $a_{j_r}^\dag$. The resulting
operator will be denoted by $X\fapp{j_0,\ldots,j_{p-1}}$.  For example, if
$X=\ket{1,0}\bra{0,1}=a_0^\dag a_1$ then $X\fapp{j,k}=a_j^\dag
a_k$. This operator acts as follows:
\[
\begin{array}{rcl}
a_j^\dag a_k\,\ket{\ldots,0,\occup_{j+1},\ldots,\occup_{k-1},0,\ldots}
&=& 0, \\
a_j^\dag a_k\,\ket{\ldots,0,\occup_{j+1},\ldots,\occup_{k-1},1,\ldots}
&=& \bigl(-1\bigr)^{\sum_{s=j+1}^{k-1}\occup_s}\,
\ket{\ldots,1,\occup_{j+1},\ldots,\occup_{k-1},0,\ldots}, \\
a_j^\dag a_k\,\ket{\ldots,1,\occup_{j+1},\ldots,\occup_{k-1},0,\ldots}
&=& 0,\\
a_j^\dag a_k\,\ket{\ldots,1,\occup_{j+1},\ldots,\occup_{k-1},1,\ldots}
&=& 0.
\end{array}
\]
Not only $X\fapp{j,k}\not=X\qapp{j,k}$ but also $X\fapp{j,k}$ is non-local in
terms of qubits: it involves all the qubits with numbers from $j$ to $k$.

A unitary {\em qubit gate} ({\em LFM, or fermionic gate}) is a unitary
operator $U$ meant to be applied to a number of qubits (LFMs); a $p$-qubit or
a $p$-LFM gate acts on the standard Hilbert space $\qb^{\otimes p}$. Operators
of the form $U\qapp{j,k}$ or $U\fapp{j,k}$ are called {\em gate
applications}. We will usually consider unitary gates up to overall phase
factors. A set of gates is also called a {\em basis}. A {\em circuit of size
$s$ in a basis $\calA$} is a composition of $s$ applications of gates from
$\calA$, i.e.\ an expression of the form
$U_s\fapp{j_{s,0},\ldots,j_{s,\,p_s-1}} \,\cdots\,
U_1\fapp{j_{1,0},\ldots,j_{1,\,p_1-1}}$, where $U_k\in\calA$. Such an
expression is evaluated by a unitary operator which is said to be {\em
represented} by the circuit.

Note that $X\fapp{j}=X\qapp{j}$, so one-LFM gates are simply parity-preserving
one-qubit gates. Up to an overall phase, such gates have the form
$\Lambda(e^{i\phi})$, where $\Lambda(U)$ denotes the controlled $U$.
(If $U$ acts on $p$ qubits then $\Lambda(U)$ acts on $p+1$
qubits; in our case $p=0$).

More generally,
$X\fapp{j,j\!+\!1,\ldots,j\!+\!p\!-\!1}
=X\qapp{j,j\!+\!1,\ldots,j\!+\!p\!-\!1}$.
This allows one to represent fermionic gates in terms of qubit gates and vice
versa. We will now show how to do that in the case $p=2$.

Suppose we want to execute a two-LFM operator $X\fapp{j,k}$ (w.\,l.\,.o.\,g.\
$j<k$). First, we move the $k$-th qubit next to the $j$-th one by swapping it
with its nearest neighbors. Then we apply
$X\fapp{j,j\!+\!1}=X\qapp{j,j\!+\!1}$ and move the $k$-th qubit back to its
original position. However, what we actually need here is to interchange LFMs,
not qubits. This is different even if the LFMs (qubits) are next to each
other!

A swap between two qubits (with numbers $0$ and $1$) is defined in the obvious
way: $(\qswap)\,:\,\ket{\occup_0,\occup_1}\mapsto\ket{\occup_1,\occup_0}$.
A swap between two LFMs is a unitary operator $(\fswap)$ such that
\begin{equation}
(\fswap)\,a_0\,(\fswap)^\dag\, =\, a_1,\qquad\quad
(\fswap)\,a_1\,(\fswap)^\dag\, =\, a_0.
\end{equation}
These equations have a unique solution (up to an overall phase factor):
\begin{equation}
(\fswap)\, =\,
I-a_0^\dag a_0-a_1^\dag a_1+a_1^\dag a_0+a_0^\dag a_1 \,=\,
\exp\Bigl(i\frac{\pi}{2}(a_0^\dag-a_1^\dag)(a_0-a_1)\Bigr)\, :\,
\left\{ \begin{array}{l}
\ket{0,0} \mapsto \ket{0,0}\\
\ket{0,1} \mapsto \ket{1,0}\\
\ket{1,0} \mapsto \ket{0,1}\\
\ket{1,1} \mapsto -\ket{1,1} .
\end{array} \right.
\end{equation}
This differs from the qubit swap by the ``-'' sign. So,
\begin{equation}
(\fswap)\ =\ (\qswap)\, D, 
\end{equation}
where $D=\Lambda(\sigma^z):\,\ket{a,b}\mapsto(-1)^{ab}\ket{a,b}$ is a ``swap
defect'' operator.

To perform the procedure described above, we do not have to actually
interchange LFMs or qubits. Instead, we can simply apply operators
$D\fapp{l,r}=D\qapp{l,r}$ to all pairs of qubits that otherwise would be
interchanged. So, any two-LFM operator $X\fapp{j,k}$\, ($j<k$) can be
represented by a qubit circuit as follows:
\begin{equation}
X\fapp{j,k}\, =\, 
D\qapp{k\!-\!1,k}\cdots D\qapp{j\!+\!1,k}\, X\qapp{j,k}\,
D\qapp{j\!+\!1,k}\cdots D\qapp{k\!-\!1,k}.
\end{equation}
Conversely, any parity-preserving two-qubit operator is represented by
a fermionic circuit:
\begin{equation}
X\qapp{j,k}\, =\, 
D\fapp{k\!-\!1,k}\cdots D\fapp{j\!+\!1,k}\, X\fapp{j,k}\,
D\fapp{j\!+\!1,k}\cdots D\fapp{k\!-\!1,k}.
\end{equation}
This method also works for operators which act on more than two LFMs (qubits).

\section{A universal set of LFM gates.}

We have shown that fermionic gates are equivalent to parity-preserving qubit
gates modulo the swap defect operator $D$. So, to obtain a universal set of
fermionic gates, we only need to find a universal set of parity-preserving
qubit gates. It will be possible to represent the operator $D$ by these gates
too, exactly or approximately. (If only approximate representation is
possible, one may have to pay extra cost when simulating a single fermionic
gate because the more qubit gates are used, the more accurate they should
be. We will avoid this problem, though).

We claim that the following gates are sufficient to represent any
parity-preserving operator to any given precision (using ancillas):
\begin{equation}\label{ppbasis}
\Lambda(e^{i\pi/4}),\qquad \Lambda(\sigma^z),\qquad
\ppext{H}:\
\ket{a,b}\mapsto \frac{1}{\sqrt{2}}\sum_{c}(-1)^{bc}\ket{a\!+\!b\!+\!c,c}.
\end{equation}
(Here $a,b,c\in{\bf F}_2=\{0,1\}$, so the expression $a+b+c$ is considered
modulo $2$). Note that $D=\Lambda(\sigma^z)$ belongs to this set of gates.

Any parity-preserving operator $U$ can be considered as a pair of operators
$(U_0,U_1)$, where $U_0$ acts on the even sector $\calH_0$ whereas $U_1$ acts
on the odd sector $\calH_1$. We will first show how to get a given $U_0$ while
not caring about $U_1$ --- we will actually get $U_1=U_0$. (The comparison
between $U_0$ and $U_1$ is made through identifying $\calH_0$ and $\calH_1$ by
the map $\sigma^x\qapp{0}$). Then we will implement operators of the form
$(I,Y)$ which can be used to correct the first step.

Any operator $X$ on $m-1$ qubits can be transformed to a parity-preserving
operator $\ppext{X}$ on $m$ qubits by using the extra qubit to maintain the
parity:
\begin{equation}\label{ppext}
\ppext{X} =\, V^{-1}\,(I_{\qb}\otimes X)\,V,\qquad\
V\,:\ \ket{\occup_0,\occup_1,\ldots,\occup_{m-1}}\mapsto
\ket{\occup_0\!+\!\ldots\!+\!\occup_{m-1},\,\occup_1,\ldots,\occup_{m-1}}.
\end{equation}
Note that the gate $\ppext{H}$ (see eq.~(\ref{ppbasis})) corresponds to the
Hadamard gate
$H=\frac{1}{\sqrt{2}}\left(\begin{array}{rr}1&1\\1&-1\end{array}\right)\,$. If
$X$ already preserves the parity, as it is the case with the operators
$\Lambda(e^{i\pi/4})$ and $\Lambda(\sigma^z)$, then $\ppext{X}=I_{\qb}\otimes
X$. (This equality is actually a characteristic property of parity-preserving
operators).

The operator $V$ is unitary. It maps the even sector $\calH_0$ onto the
subspace $\calL_0$ which consists of vectors $\ket{0}\otimes\ket{\xi}$\,
($\ket{\xi}\in\qb^{\otimes(m-1)}$). Any operator on this subspace extends to
an operator of the form $I_{\qb}\otimes X$. Hence any $U_0\in\LL(\calH_0)$
extends to an operator of the form $\ppext{X}$. Note that $V$ and
$I_{\qb}\otimes X$ commute with $\sigma^x\qapp{0}$, so $\ppext{X}$ also
commutes with $\sigma^x\qapp{0}$. Therefore $\ppext{X}=(U_0,U_1)$, where
$U_1=\sigma^x\qapp{0}\,U_0\,\sigma^x\qapp{0}$, or simply $U_1=U_0$ if the
identification between $\calH_0$ and $\calH_1$ is used.

If an operator $X$ is represented by a quantum circuit $A_L\cdots A_1$, one
can replace each gate application $A_k$ by $\ppext{A_k}$ to get a quantum
circuit for $\ppext{X}$. Indeed, eq.~(\ref{ppext}) defines a $*$-algebra
homomorphism, i.e.
\[
\ppext{X_1{+}X_2}=\ppext{X_1}+\ppext{X_2},\quad\
\ppext{cX}=c\ppext{X},\quad\
\ppext{X_1X_2}=\ppext{X_1}\ppext{X_2},\quad\
\ppext{X^\dag}={\ppext{X}\,}^\dag,\quad\
\ppext{I}=I.
\]
It follows that any universal gate set $\calA$ transforms to a set of
parity-preserving gates $\ppext{\calA}$ which are universal on the even
sector.  The following gate set is known to be universal~\cite{universality}:
$\calA=\Bigl\{\Lambda(e^{i\pi/4}),\,\Lambda(\sigma^z),\,H\Bigr\}$. The
corresponding gate set $\ppext{\calA}$ is given by eq.~(\ref{ppbasis}). (The
parity-preserving gates $\Lambda(e^{i\pi/4})$ and $\Lambda(\sigma^z)$ are
copied from $\calA$ to $\ppext{\calA}$ unchanged). Thus the
basis~(\ref{ppbasis}) allows one to obtain at least unitary gates of the form
$(U_0,U_0)$. 

With parity-preserving operators, we can still use one of the standard
techniques in quantum circuit design --- gates with quantum control. Indeed,
if $U$ preserves the parity then $\Lambda(U)$ also does. More generally,
$\ppext{\Lambda(X)}=\Lambda(\ppext{X})$ (up to a permutation of the control
qubit and the parity qubit). So, if we can represent $\Lambda(X)$ by a
circuit,\footnote{Implementing $\Lambda(X)$ in the basis $\calA$ is only
slightly harder than implementing $X$. Indeed, for each gate $X$ from the
basis $\calA$, the operator $\Lambda(X)$ can be represented by a circuit in
the same basis {\em exactly}\/. Therefore, the circuit for $\Lambda(X)$ will
be larger than the circuit for $X$ only by a constant factor. Note, however,
that $X$ may have been implemented up to a phase factor; this phase factor
becomes important when we consider $\Lambda(X)$. The necessary correction can
be achieved by an operator $\Lambda(e^{i\phi})$ which should be implemented
separately.} we can also obtain a circuit for $\Lambda(\ppext{X})$ by the
procedure described above. For example, the operator $\Lambda(\sigma^x)[1,2]$
can be represented as $H\qapp{2}\,\Lambda(\sigma^z)\qapp{1,2}\,H\qapp{2}$,
hence
\[
\Lambda(\ppext{\sigma^x})\qapp{1,0,2}\,=\
\ppext{H}\qapp{0,2}\ \Lambda(\sigma^z)\qapp{1,2}\ \ppext{H}\qapp{0,2}.
\]
Here we use qubit $0$ to maintain the parity whereas $1$ is the control
qubit. (The notation $\Lambda(\ppext{X})\qapp{\ldots}$ suggests that the
control qubit goes first).

Now we are in a position to implement operators of the form $(I,Y)$ using the
gates~(\ref{ppbasis}). We will also use one ancilla which will be assigned the
number $m$. First, we execute the operator
\begin{equation}
\begin{array}{l}
W:\ \ket{\occup_0,\occup_1,\ldots,\occup_{m-1},\occup_m}\,\mapsto\,
\ket{\occup_0\!+\!\ldots\!+\!\occup_{m-1},\,\occup_1,\ldots,\occup_{m-1},\,
{\occup_1\!+\!\ldots\!+\!\occup_m}}, \medskip\\
W\, =\ \Lambda(\ppext{\sigma^x})\qapp{m\!-\!1,0,m}\ \cdots\
\Lambda(\ppext{\sigma^x})\qapp{1,0,m},
\end{array}
\end{equation}
where
$\Lambda(\ppext{\sigma^x}):\,\ket{a,b,c}\mapsto\ket{a,b\!+\!a,c\!+\!a}$. Now
qubit $0$ indicates the total parity.

Let $\ppext{X}=(Y,Y)$. Since $\ppext{X}$ can be represented according
to~(\ref{ppext}), and because
\[
W\,=\,V^{-1}\qapp{m,1,\ldots,m\!-\!1}\ V,\quad\ 
\mbox{where}\ V=V\qapp{0,\ldots,m\!-\!1},
\]
we conclude that
\begin{equation}
W^{-1}\, \Lambda(\ppext{X})\qapp{0,m,1,\ldots,m\!-\!1}\ W\ =\ 
V^{-1}\, \Lambda(X)\qapp{0,1,\ldots,m\!-\!1}\ V\ =\ 
U\otimes I_\qb,\quad\ U=(I,Y).
\end{equation}
Note that in this case, the ancilla is not affected no matter what its initial
state was; one can even use a data qubit as the ancilla.

We have proved that the gate set~(\ref{ppbasis}) is universal in the class of
parity-preserving operators. It remains to represent these gates in terms of
creation and annihilation operators. The first two are simple:
\begin{equation}\label{fbasis1}
\Lambda(e^{i\pi/4})=\exp\Bigl(i\frac{\pi}{4}a_0^\dag a_0\Bigr),\qquad
\Lambda(\sigma^z)=\exp(i\pi a_0^\dag a_0a_1^\dag a_1).
\end{equation}
Unfortunately, the gate $\ppext{H}$ in the fermionic representation looks
ugly, so we first represent it as follows:
\[
\ppext{H}\qapp{0,1}\,=\
\Lambda(-i)\qapp{1}\ \ppext{G}\qapp{0,1}\ \Lambda(-i)\qapp{1},\qquad\
G=\left(\begin{array}{cc} 1 & i \\ i & 1 \end{array}\right) .
\]
So, the fermionic gates~(\ref{fbasis1}), together with
\begin{equation}\label{ppextG}
\ppext{G}\,=\,
\exp\Bigl(-i\frac{\pi}{4}(a_0-a_0^\dag)(a_1+a_1^\dag)\Bigr) \,=\,
\exp\Bigl(i\frac{\pi}{4}(a_0^\dag a_1+a_1^\dag a_0)\Bigr)
\exp\Bigl(i\frac{\pi}{4}(a_1 a_0+a_0^\dag a_1^\dag)\Bigr) ,
\end{equation}
form a universal set. Obviously, this gate set is also universal:
\begin{equation}\label{fbasis}
\left\{\quad
\begin{array}{l}
\exp\Bigl(i\frac{\pi}{4}a_0^\dag a_0\Bigr),
\medskip\\
\exp\Bigl(i\frac{\pi}{4}(a_0^\dag a_1+a_1^\dag a_0)\Bigr), 
\medskip\\
\exp\Bigl(i\frac{\pi}{4}(a_1 a_0+a_0^\dag a_1^\dag)\Bigr),
\medskip\\
\exp(i\pi a_0^\dag a_0a_1^\dag a_1)
\end{array}
\quad\right\}.
\end{equation}

\section{Fast simulation procedures}\label{sec_simulation}

So far we have been using the standard identification~(\ref{fvq}) between the
Foch space $\calH$ and the Hilbert space of $m$ qubits $\qb^{\otimes m}$. This
identification has allowed us to consider qubit gate applications
$X\qapp{j,k}\in\LL(\qb^{\otimes m})$ and LFM gate applications
$X\fapp{j,k}\in\calF\subseteq\LL(\calH)$ as operators acting on the same
space. Now we are going to discuss a more general way of simulating LFMs by
qubits (or vice versa). Let $J$ be an {\em encoding} of $m$ LFMs by $m'$
qubits, i.e.\ $J:\calH\to\qb^{\otimes m'}$ is a unitary embedding. ($J$ being
a unitary embedding means that $J^\dag J=I_\calH$. Note that $JJ^\dag$
is the projector onto $\calL=\Image J\subseteq\qb^{\otimes m'}$). We say that
an operator $U'\in\LL(\qb^{\otimes m'})$ {\em represents} an operator
$U\in\LL(\calH)$ if
\begin{equation}
\begin{array}{ccc}
\calH & \stackrel{U}{\longleftarrow} & \calH \\
\downarrow\lefteqn{\scriptstyle J} && \downarrow\lefteqn{\scriptstyle J}\\
\qb^{\otimes m'} & \stackrel{U'}{\longleftarrow} & \qb^{\otimes m'}
\end{array}
\ \mbox{commutes}, \qquad \mbox{i.e.}\quad JU=U'J.
\end{equation}
The operator $U$, as well as the LFMs or qubits equivalent to them, are called
{\em logical}. The $m'$ qubits the operator $U'$ acts on are called {\em code
qubits}.

In this section we will show that each LFM gate can be simulated by $O(\log
m)$ qubit gates. Surprisingly, this does not require quantum codes in the
proper sense, i.e.\ $J:\calH\to\qb^{\otimes m}$ is a map onto. Moreover, $J$
takes basis vectors to basis vectors. We will also show how to simulate qubit
gates by LFM gates at a constant cost using a subspace of the Foch space.

Gate action on a set of qubits (LFMs) can be described as follows. First, we
{\em extract} two (or some other number) qubits or LFMs from the {\em quantum
memory} which is now considered as a ``black box''. We place these qubits at
positions $-2$ and $-1$. Then we apply the gate and put the qubits (LFMs) back
into the memory. In this description, it does not matter what gate we apply, a
qubit one or a fermionic one (because $X\fapp{-2,-1}=X\qapp{-2,-1}\,$). Once
extracted, LFMs can be regarded as qubits --- the difference is in the way we
extract them. With qubits, we apply the operator
\[
\qext{j}\,:\ \qb^{\otimes m}\to\qb\otimes\qb^{\otimes m}\ :\
\ket{\occup_0}\otimes\cdots\otimes\ket{\occup_j}\otimes\cdots\otimes
\ket{\occup_{m-1}}\,\mapsto\,
\ket{\occup_j}\,\otimes\,\ket{\occup_0}\otimes\cdots\otimes
\ket{0}\otimes\cdots\otimes\ket{\occup_{m-1}}.
\]
This is a unitary embedding into a larger Hilbert space; it can be represented
as adding an ancilla in the state $\ket{0}$ followed by a unitary
operator. With fermions, we move the $j$-th LFM through all the LFMs left to
it, so we should take swap defects into account. Thus we get another unitary
embedding:
\begin{equation}
\begin{array}{l}
\fext{j}\,:\ \calH\to\qb\otimes\calH, \medskip\\
\fext{j}\,
\ket{\occup_0,\ldots,\occup_j,\ldots,\occup_{m-1}}\ =\
\bigl(-1\bigr)^{\occup_j\sum_{s=0}^{j-1}\occup_s}\,
\ket{\occup_j}\otimes\ket{\occup_0,\ldots,0,\ldots,\occup_{m-1}}.
\end{array}
\end{equation}
It is easy to verify that our recipe is correct, i.e.\ applying the operator
$X\qapp{-2,-1}$ after extracting two LFMs is equivalent to applying
$X\fapp{j,k}$ before the extraction:
\begin{equation}
X\qapp{-2,-1}\,\Bigl(I_\qb\otimes\fext{k}\Bigr)\,\fext{j} \ =\
\Bigl(I_\qb\otimes\fext{k}\Bigr)\,\fext{j}\,X\fapp{j,k}.
\end{equation}
(Our notations are somewhat confusing so this comment could be helpful.
When we extract the $j$-th LFM by
applying $\fext{j}$, we add one qubit at position $-1$. When we then extract
the $k$-th LFM, this qubit moves to position $-2$ while another one is being
inserted; this can be described by the operator $I_\qb\otimes\fext{k}$.)

Suppose we want to simulate LFMs by qubits at low cost. The problem with the
standard encoding~(\ref{fvq}) is that multiplication by the factor
$\bigl(-1\bigr)^{\occup_j\sum_{s=0}^{j-1}\occup_s}$ requires too many
operations. The simplest solution would be to store
$y_j=\sum_{s=0}^{j-1}\occup_s$ instead of $\occup_j$. (Remember that we
consider $\occup_s$ as residues$\pmod{2}$, so the sum is also
taken$\pmod{2}\,$). This does solve the problem but also creates a new one:
when $\occup_j$ becomes $0$ as a result of extraction, we have to modify all
$y_k$,\, $k>j$.  So, we need to balance the complexity of computing
$\sum_{s=0}^{j-1}\occup_s$ and that of updating the encoded quantum memory
when $\occup_j$ changes. This kind of trade-off can be achieved by storing
some partial sums $\sum_{s=a}^{b}\occup_s$.

In general, we will use encodings of the form
\[
\begin{array}{l}
\displaystyle
J\,:\ \calH\to\qb^{\otimes m}\ :\
\ket{\occup_0,\ldots,\occup_{m-1}}\mapsto
\ket{x_0}\otimes\cdots\otimes\ket{x_{m-1}},\medskip\\
\displaystyle
x_j=\sum_{s\in S(j)}\occup_s,\quad\
S(j)\subseteq\{0,\ldots,m-1\}.
\end{array}
\]
We start with an example of such an encoding for $m=8$
(the diagram next to the equation illustrates grouping of $n_s$ into $x_j$):
\[
\begin{array}{llll}
x_0=\occup_0\qquad & x_2=\occup_2\qquad & 
x_4=\occup_4\qquad & x_6=\occup_6 \medskip\\
\multicolumn{2}{l}{x_1=\occup_0+\occup_1} &
\multicolumn{2}{l}{x_5=\occup_4+\occup_5} \medskip\\
\multicolumn{4}{l}{x_3=\occup_0+\occup_1+\occup_2+\occup_3} \medskip\\
\multicolumn{4}{l}{x_7=
\occup_0+\occup_1+\occup_2+\occup_3+\occup_4+\occup_5+\occup_6+\occup_7}
\end{array}
\qquad
\fbox{$
\fbox{$\fbox{$\fbox{$\occup_0$}\ \occup_1$}\ \fbox{$\occup_2$}\ \occup_3$}\
\fbox{$\fbox{$\occup_4$}\ \occup_5$}\ \fbox{$\occup_6$}\ \occup_7$}
\]
A binary tree structure is apparent here. To proceed, we will represent the
qubit indices $0,\ldots,m-1$ by binary strings. The length of these strings is
not fixed; we may add an arbitrary number of zeros to the beginning of a
string, e.g.\ $3=\binary{11}=\binary{011}=\binary{0011}$.

Let us define a partial order $\preceq$ on the set of binary strings. We write
$\binary{\alpha_{t-1}\ldots\alpha_0}\preceq\binary{\beta_{t-1}\ldots\beta_0}$
if $\alpha_l=\beta_l$ for $l\ge l_0$ while $\beta_{l_0-1}=\ldots=\beta_0=1$
(for some $l_0$). For example,
\[
\left.\begin{array}{r}
\left.\begin{array}{r}
\binary{000}\prec\binary{001}\medskip\\
\binary{010}
\end{array}\right\} \prec\binary{011}\medskip\\
\binary{100}\prec\binary{101}\medskip\\
\binary{110}
\end{array}\right\} \prec\binary{111},
\]
where $j\prec k$ means that $j\preceq k$ but $j\not=k$. Note that if $j\prec
k$ then $j<k$.

Now we can specify our encoding for arbitrary $m$:
\begin{equation}\label{enc1}
\ket{\occup_0,\ldots,\occup_{m-1}}\mapsto
\ket{x_0}\otimes\cdots\otimes\ket{x_{m-1}},\qquad\
x_j=\sum_{s\preceq j}\occup_s.
\end{equation}
It is important that each $\occup_s$ enters only $O(\log m)$ of $x_j$. Hence
applying $\sigma^x\qapp{s}$ to one logical qubit amounts to logarithmically
many $\sigma^x$ gates being applied to the code qubits.

The inverse transformation (from $x_j$ to $\occup_s$) is also simple:
\begin{equation}\label{enc1inv}
\begin{array}{l}
\multicolumn{1}{c}{\displaystyle\occup_j=x_j-\sum_{s\in K(j)}x_s,}
\medskip\\
\binary{\alpha_{t-1}\ldots\alpha_0}\in
K\Bigl(\binary{\beta_{t-1}\ldots\beta_0}\Bigr)
\quad\mbox{if and only if} \medskip\\
\qquad
\beta_{l_0}=\ldots=\beta_0=1,\quad
\alpha_l=\beta_l\ \mbox{for}\ l\not=l_0,\,\ \alpha_{l_0}=0\quad
\mbox{(for some $l_0$)}.
\end{array}
\end{equation}
The sum in~(\ref{enc1inv}) contains only $O(\log m)$ terms. Moreover,
$y_j=\sum_{s=0}^{j-1}\occup_s$ can be also expressed as a sum of $O(\log m)$
numbers $x_s$:
\begin{equation}\label{enc1sum}
\begin{array}{l}
\multicolumn{1}{c}{\displaystyle y_j=\sum_{s\in L(j)}x_s,}
\medskip\\
\binary{\alpha_{t-1}\ldots\alpha_0}\in
L\Bigl(\binary{\beta_{t-1}\ldots\beta_0}\Bigr)\quad\ \mbox{if and only if}
\medskip\\
\qquad
\beta_{l_0}=1,\quad
\alpha_l=\beta_l\ \mbox{for}\ l>l_0,\,\  \alpha_{l_0}=0,\,\
\alpha_{l_0-1}=\ldots=\alpha_0=1\quad \mbox{(for some $l_0$)}.
\end{array}
\end{equation}

It remains to actually represent the LFM extraction operator $\fext{j}$ by
acting on the code qubits. Since $\fext{j}$ increases the number of qubits by
$1$, we should add this extra qubit first. We place it at position $-1$ and
initialize by $\ket{0}$. Then we make its value equal to $\occup_j$ by
applying the operators $\Lambda(\sigma^x)\qapp{s,-1}$,\,
$s\in K(j)\cup\{j\}$ (see eq.~(\ref{enc1inv})). After that, $\occup_j$ can be
turned into $0$ by executing the operators $\Lambda(\sigma^x)\qapp{-1,k}$,\,
$k\succeq j$ (see eq.~(\ref{enc1})). Finally, we multiply by
$(-1)^{\occup_jy_j}$ by applying $\Lambda(\sigma^z)\qapp{-1,s}$,\, $s\in L(j)$
(see eq.~(\ref{enc1sum})). To summarize, we execute the operator
\begin{equation}
U'\,=\ \prod_{s\in L(j)}\! \Lambda(\sigma^z)\qapp{-1,s}\
\prod_{k\succeq j} \Lambda(\sigma^x)\qapp{-1,k}
\prod_{s\in K(j)\cup\{j\}}\!\! \Lambda(\sigma^x)\qapp{s,-1}
\end{equation}
which has the property\,
$(I_\qb\otimes J)\, \fext{j}\, \ket{\xi}\,=\,
U'\Bigl(\ket{0}\otimes J\ket{\xi}\Bigr)$\, for any $\ket{\xi}\in\calH$.\,
This requires $O(\log m)$ operations.
\smallskip

Simulating qubits by LFMs is easier and faster. One can use this simple
encoding:
\begin{equation}\label{enc2}
\ket{\occup_0}\otimes\ket{\occup_1}\otimes\cdots\otimes\ket{\occup_{m-1}}\,
\mapsto\,
\ket{\occup_0,\occup_0,\,\occup_1,\occup_1,\ldots,\occup_{m-1},\occup_{m-1}},
\end{equation}
i.e. each qubit is represented by a pair of LFMs with an even number of
fermions,\, $\ket{0}\mapsto\ket{00}$,\,\, $\ket{1}\mapsto\ket{11}$. Each
two-qubit operator $X$ is represented by a four-qubit operator $X'$. The
operators $X'\qapp{2j,2j\!+\!1,2k,2k\!+\!1}$ and
$X'\fapp{2j,2j\!+\!1,2k,2k\!+\!1}$ act the same way on the code
subspace. So, when we use the encoding~(\ref{enc2}), it does not matter
whether the code elements are qubits or LFMs. The simulation cost is just
a constant.

\section{Majorana fermions}

It is possible, at least mathematically, to split each local fermionic mode
into two objects. These halves of LFMs are called Majorana fermions.\footnote{
In field theory, this term usually means something more specific, but it is
sometimes used in this sense too.}

Let us introduce a set of Hermitian operators $c_j$\, ($j=0,\ldots,2m-1$):
\begin{equation} \label{MFdef}
c_{2k}\,=\,a_k+a_k^\dag
\,=\,\sigma^x\qapp{k} \prod_{j=0}^{k-1}\sigma^z\qapp{j}\,, \qquad\quad
c_{2k+1}\,=\,\frac{a_k-a_k^\dag}{i}
\,=\,\sigma^y\qapp{k} \prod_{j=0}^{k-1}\sigma^z\qapp{j}.
\end{equation}
These operators satisfy the commutation relations
\begin{equation}
c_j c_k+c_k c_j=2\delta_{jk}.
\end{equation}
We can define new locality rules on the algebra
$\calF=\LL(\calH_0)\oplus\LL(\calH_1)$ which will be now denoted by $\calM$.
(Also $\bar{\calF}=\LL(\calH)$ will be denoted by $\bar{\calM}$). We say that
there are $2m$ sites called {\em Majorana fermions}. For each set of sites
$S\subseteq\{0,\ldots,2m-1\}$, let $\bar{\calM}(S)\subseteq\bar{\calM}$ be the
subalgebra generated by $c_j$,\, $j\in S$. (Such an algebra is known as {\em
complex Clifford algebra}). Then $\calM(S)=\bar{\calM}(S)\cap\calM$, i.\,e.\
{\em physical operators} on $S$ are linear combinations of even products of
$c_j$,\, $j\in S$.

According to this definition, Majorana fermions ``exist'' in any Fermi system.
A nontrivial thing is that it is possible (at least theoretically) to pair
them up by interaction, so that few Majorana fermions remain unpaired and {\em
separated from each other}\/~\cite{q-wire}. Such systems could be used as
decoherence-free quantum memory. Indeed, a single Majorana fermion can not
interact with the environment by itself (because the operator $c_j$ is not
physical), so the decoherence can only arise from environment-mediated
interaction of two Majorana fermions. But if they are well separated in space,
such interaction should be exponentially small (a finite correlation length in
the environment is assumed). Roughly speaking, we keep two halves of a qubit
apart, so the qubit is decoherence-free!

An application of a {\em Majorana gate} is defined as in the case of LFMs.
Let $X$ be a physical operator acting on $p$ Majorana fermions numbered $0$
through $p-1$. Then $X$ can be expanded into (even) products of $c_j$,\,
$j=0,\ldots,p-1$. If we substitute $c_{j_r}$ for $c_r$, we will get an
operator $X\mapp{j_0,\ldots,j_{p-1}}\in\calM(\{j_0,\ldots,j_{p-1}\})$.

To find a universal set of Majorana gates, it suffice to rewrite the
operators~(\ref{fbasis1}), (\ref{ppextG}) in terms of $c_0=a_0+a_0^\dag$,\,
$c_1=-i(a_0-a_0^\dag)$,\, $c_2=a_1+a_1^\dag$,\,
$c_3=-i(a_1-a_1^\dag)$.
\[
\begin{array}{c}
a_0^\dag a_0 = \frac{1}{2}(1+ic_0c_1),\qquad\ 
a_1^\dag a_1 = \frac{1}{2}(1+ic_2c_3), \medskip\\
\exp\Bigl(i\frac{\pi}{4}a_0^\dag a_0\Bigr) =
e^{i\pi/8} \exp\Bigl(-\frac{\pi}{8}c_0c_1\Bigr),\qquad\quad
\exp\Bigl(-i\frac{\pi}{4}(a_0-a_0^\dag)(a_1+a_1^\dag)\Bigr)=
\exp\Bigl(\frac{\pi}{4}c_1c_2), \medskip\\
\exp\Bigl(i\pi a_0^\dag a_0 a_1^\dag a_1\Bigr) =
e^{-i\pi/4}
\exp\Bigl(\frac{\pi}{4}c_0c_1\Bigr) \exp\Bigl(\frac{\pi}{4}c_2c_3\Bigr)
\exp\Bigl(i\frac{\pi}{4}c_0c_1c_2c_3\Bigr).
\end{array}
\]
Hence the following gate set is universal (up to phase factors):
\begin{equation}\label{mbasis}
\left\{\
\exp\Bigl(\frac{\pi}{8}c_0c_1\Bigr),\,\
\exp\Bigl(i\frac{\pi}{4}c_0c_1c_2c_3\Bigr)
\ \right\}.
\end{equation}
\smallskip
 
The second gate in the set~(\ref{mbasis}) describes four-particle
interaction. When it comes to physical implementation, this gate will be
particularly difficult to realize. Unfortunately, it is not possible to do
universal quantum computation by acting on three or fewer Majorana fermions at
a time. This way one can only generate the group of unitary operators which
act by conjugation as follows:
\begin{equation}\label{lintrans}
Uc_{j}U^\dag \,=\, \sum_{k}\beta_{jk}c_k,
\end{equation}
where $\beta\in\mathop{SO}(m)$, i.e.\ $(\beta_{jk})$ is a real orthogonal
matrix with determinant $+1$.

\section{An alternative to the four-particle Majorana gate}

In this section we show that, for the purpose of universal computation with
Majorana fermions, the gate $\exp(i\frac{\pi}{4}c_0c_1c_2c_3)$ can be replaced
by a nondestructive eigenvalue measurement of the operator $c_0 c_1 c_2
c_3$. (A measurement being nondestructive means that vectors in each of the
eigenspaces remain intact; in other words, no extra information is learned or
leaks to the environment). Such measurements might be easier to implement,
they are also useful in some theoretical application of fermionic
computation~\cite{Ising}.

Let us assume that the following operations are possible:
\begin{enumerate}
\item Applying the unitary gate $R=\exp\Bigl(\frac{\pi}{4}c_0c_1\Bigr)$.
(Note that $Rc_0R^\dag=-c_1$,\, $Rc_1R^\dag=c_0$).
\item Creation of an ancilla pair in a state which is the eigenstate of
$c_{2k}c_{2k+1}=i(1-2a_k^\dag a_k)$ corresponding to the eigenvalue $i$.
\item Eigenvalue measurement of $c_jc_k$.
\item Nondestructive eigenvalue measurement of $c_jc_kc_rc_s$.
\end{enumerate}
Moreover, we can base the choice of the next operation on the previous
measurement outcomes. (One may call such quantum computation {\em adaptive}).
Of course, the amount of classical computation involved in this choice should
not be too large, so we should better include it into the overall size of the
quantum circuit.\smallskip

\noindent{\em Remark.} In the standard scheme of quantum computation,
measurements in the middle of computation are redundant and can be simulated
by unitary gates (e.\,g.\ see~\cite{AKN}). However, this is only true if
one uses a universal set of unitary gates. Otherwise measurements and
adaptiveness can indeed add extra power to unitary operators.\smallskip

Suppose we want to apply the operator $\exp(i\frac{\pi}{4}c_0c_1c_2c_3)$.
Let Majorana fermions 4 and 5 form an ancilla pair, so the input state of
the system satisfies
\begin{equation}\label{initcond}
(c_4+ic_5)\ket{\Psi_{\rm in}}=0.
\end{equation}
We measure the eigenvalue of $c_0c_1c_3c_4$. The outcome is either $+1$ or
$-1$, which means that $\ket{\Psi_{\rm in}}$ gets multiplied by the projector
$\Pi^{(4)}_{+1}=\frac{1}{2}(1+c_0c_1c_3c_4)$ or
$\Pi^{(4)}_{-1}=\frac{1}{2}(1-c_0c_1c_3c_4)$, respectively. (More exactly,
$\ket{\Psi_{\rm in}}\mapsto
p_{\pm 1}^{-1/2}\Pi^{(4)}_{\pm 1}\ket{\Psi_{\rm in}}$,
where $p_{z}$ is the probability to get outcome $z$). Then we measure the
eigenvalue of $c_2c_4$. The possible eigenvalues $+i$ and $-i$ correspond to
the projectors $\Pi^{(2)}_{+i}=\frac{1}{2}(1-ic_2c_4)$ and
$\Pi^{(2)}_{-i}=\frac{1}{2}(1+ic_2c_4)$. We claim that after some correction
depending on the measurements outcomes, we will effectively execute the
operator $\exp(i\frac{\pi}{4}c_0c_1c_2c_3)$ while leaving the ancilla pair
intact. Indeed,
\begin{equation}
\begin{array}{rc}
\multicolumn{2}{l}{%
\exp\Bigl(i\frac{\pi}{4}c_0c_1c_2c_3\Bigr)\,\ket{\Psi_{\rm in}}\ =}
\smallskip\\
=\ 2\,\exp\Bigl(\frac{\pi}{4}c_2c_5\Bigr)\,
\frac{1}{2}\Bigl( 1-ic_2c_4\Bigr)\,
\frac{1}{2}\Bigl(1+c_0c_1c_3c_4\Bigr)\,
\ket{\Psi_{\rm in}} &=\smallskip\\
=\ 2\,i \exp\Bigl(\frac{\pi}{2}c_0c_1\Bigr)\exp\Bigl(\frac{\pi}{2}c_2c_3\Bigr)
\exp\Bigl(\frac{\pi}{4}c_2c_5\Bigr)\, \frac{1}{2}\Bigl( 1-ic_2c_4\Bigr)\,
\frac{1}{2}\Bigl(1-c_0c_1c_3c_4\Bigr)\,
\ket{\Psi_{\rm in}} &=\smallskip\\
=\ 2\,i \exp\Bigl(\frac{\pi}{2}c_0c_1\Bigr)\exp\Bigl(\frac{\pi}{2}c_2c_3\Bigr)
\exp\Bigl(-\frac{\pi}{4}c_2c_5\Bigr)\, \frac{1}{2}\Bigl(1+ic_2c_4\Bigr)\,
\frac{1}{2}\Bigl( 1+c_0c_1c_3c_4\Bigr)\,
\ket{\Psi_{\rm in}} &=\smallskip\\
=\ 2\,\exp\Bigl(-\frac{\pi}{4}c_2c_5\Bigr)\, \frac{1}{2}\Bigl(
1+ic_2c_4\Bigr)\, \frac{1}{2} \Bigl(1-c_0c_1c_3c_4\Bigr)\,
\ket{\Psi_{\rm in}} &
\end{array}
\end{equation}
(we have used eq.~(\ref{initcond})). Thus we can apply a suitable correction
operator $U_{yz}$ in each of the four cases
($U_{+i,+1}=\exp(\frac{\pi}{4}c_2c_5)$ if the outcomes were $+1$ and $+i$,
etc.) so that
\[
\exp\Bigl(i\frac{\pi}{4}c_0c_1c_2c_3\Bigr)\,\ket{\Psi_{\rm in}} \,=\,
2\,U_{yz}\Pi^{(2)}_{y}\Pi^{(4)}_{z}\,\ket{\Psi_{\rm in}}.
\]
Each of the four outcome combinations occurs with probability
$2^{-2}=\frac{1}{4}$. The final state is always the desired one,
$\ket{\Psi_{\rm fin}}=
\exp(i\frac{\pi}{4}c_0c_1c_2c_3\Bigr)\ket{\Psi_{\rm in}}$.

\section{Superfast simulation of fermions on a graph}

The results of Sec.~\ref{sec_simulation} suggest that fermions have slightly
more computational power than qubits. The logarithmic slowdown in simulation
of fermions seems to be inevitable in the general case. However, in the
physical world fermions (e.\,g.\ electrons) interact locally not only in the
sense that the interaction is pairwise, but also in the geometric sense: a
particle can not instantly jump to another position far away. Such physical
interactions might be easier to simulate. In this section we study an abstract
model of geometrically local interactions. The result is that geometrically
local gates can indeed be simulated without any substantial slowdown, i.\,e.\
the simulation cost is constant. Therefore one can speculate that, in
principle, electrons might not be fundamental particles but, rather,
excitations in a (nonperturbative) system bosons. Of course, this is only a
logical possibility which may or may not be true.

What follows is a definition of the model. Consider a {\em connected
unoriented graph}\/ $\Gamma=(M,E)$, where $M=\{0,\ldots,m-1\}$ is the set of
vertices, and $E\subset M\times M$ is the set of edges. (As the graph is
unoriented, $(j,k)$ and $(k,j)$ either both belong to $E$ or both do not
belong to $E$). We will assume that the degree of each vertex is bounded by
some constant $d$. Let us put a {\em local fermionic mode} on each {\em
vertex}. We will consider only the {\em even sector} of the system, $\calH_0$,
i.\,e.\ the total number of fermions is required to be even. The {\em allowed
unitary operations} are one-LFM gates and two-LFM gates applied to any pair of
vertices connected by an edge.

We are going to identify the Hilbert space $\calH_0$ with a codespace of a
certain symplectic (stabilizer) code~\cite{sympcodes}\footnote{The term
``stabilizer code'' has become traditional but it is somewhat confusing
because any code can be defined in terms of stabilizer operators. (It is
actually practical to do so for nonbinary codes related to
anyons~\cite{Kit1}). We prefer to use the more specific terms ``symplectic
code''~\cite{Kit2}.} so that each elementary operator of the form $a_k^\dag
a_k$, as well as $a_j^\dag a_k$, $a_k^\dag a_j$, $a_j a_k$ or
$a_k^\dag a_j^\dag$, where $(j,k)\in E$, be represented by operators
acting on $O(d)$ qubits. Then each one-LFM gate and each two-LFM gate applied
to neighboring vertices will be also represented by an operator acting on
$O(d)$ qubits. As $d={\rm const}$, this means one can simulate each of the
allowed fermionic operations by a constant number of one-qubit and two-qubit
gates.

It will be convenient to use the Majorana fermions operators $c_{2k}$,
$c_{2k+1}$ (see eq.~(\ref{MFdef})) instead of $a_k$, $a_k^\dag$. The list
of elementary operators to be represented can be reduced to these ones:
\begin{equation}
\begin{array}{rcl@{\quad}l}
B_k &=& -ic_{2k}c_{2k+1} & \mbox{for each vertex}\ k,
\smallskip\\
A_{jk} &=& -ic_{2j}c_{2k} & \mbox{for each edge}\ (j,k)\in E.
\end{array}
\end{equation}
These operators satisfy the following relations:
\begin{equation}\label{simple_relations}
B_k^\dag=B_k,\quad A_{jk}^\dag=A_{jk}, \qquad\quad
B_k^2=1,\quad  A_{jk}^2=1,\qquad\quad A_{kj}=-A_{jk},
\end{equation}
\begin{equation}\label{commutation_relation}
B_kB_l=B_lB_k,\qquad
A_{jk}B_l=(-1)^{\delta_{jl}+\delta_{kl}}B_lA_{jk},\qquad
A_{jk}A_{ls}=(-1)^{\delta_{jl}+\delta_{js}+\delta_{kl}+\delta_{ks}}
A_{ls}A_{jk},
\end{equation}
\begin{equation}\label{cycle_relation}
i^p A_{j_0,j_1}A_{j_1,j_2}\cdots A_{j_{p-2},j_{p-1}}A_{j_{p-1},j_0} \,=\, 1
\quad \mbox{for any closed path on the graph}.
\end{equation}
It is easy to prove that $B_k$, $A_{jk}$ modulo these relations generate the
algebra of physical operators $\calF=\LL(\calH_0)\oplus\LL(\calH_1)$. However,
we are considering only the even sector now. Having been restricted to
$\calH_0$, the operators $B_k$ satisfy an additional relation (which was false
in $\calF$):
\begin{equation}\label{parity_relation}
\prod_{k}B_k=1.
\end{equation}
Hence the algebra $\LL(\calH_0)$ is generated by $B_k$, $A_{jk}$
modulo the relations (\ref{simple_relations})--(\ref{parity_relation}).

To construct the code, we put a qubit on each edge of the graph. Thus
$\sigma^\alpha_{jk}=\sigma^\alpha_{kj}$ denotes the Pauli operator
$\sigma^\alpha$\, $(\alpha=x,y,z)$ acting on the edge $(j,k)$. The operators
$B_k$, $A_{jk}$ defined above will be identified with some operators
$\tilde{B}_k$, $\tilde{A}_{jk}$ acting on the code subspace $\calL$ (which
will be defined later). We start with defining the action of $\tilde{B}_k$ and
$\tilde{A}_{jk}$ on the entire Hilbert space of the qubits. Our construction
depends on two arbitrary choices. Firstly, we choose orientation for each edge
of the graph. This can be described by a matrix $(\epsilon_{jk})$ such that
$\epsilon_{kj}=-\epsilon_{jk}$,\, $\epsilon_{jk}=\pm 1$ for each edge
$(j,k)\in E$. Secondly, for each vertex $k$, we order all incident edges
$(j,k)$. This order will be denoted by $\locless{k}$. Now we put
\begin{equation}
\begin{array}{rcl}
\tilde{B}_k &=& \displaystyle \prod_{j:\,(j,k)\in E}\sigma^z_{jk},
\medskip\\
\tilde{A}_{jk} &=& \displaystyle \epsilon_{jk}\,\sigma^x_{jk}\,
\prod_{l:\,(l,j)\locless{j}(k,j)}\!\!\sigma^z_{lj}\,
\prod_{s:\,(s,k)\locless{k}(j,k)}\!\!\sigma^z_{sk}.
\end{array}
\end{equation}
These operators satisfy the relations analogous to
(\ref{simple_relations}), (\ref{commutation_relation}) and
(\ref{parity_relation}), but not (\ref{cycle_relation}).

Finally, we define the code subspace $\calL\subseteq\qb^{\otimes u}$ (where
$u$ is the number of qubits) by imposing stabilizer conditions:
\begin{equation}\label{stabcond}
\begin{array}{c}
\ket{\psi}\in\calL\quad \mbox{if and only if}\quad
\tilde{C}_{j_0,\ldots,j_{p-1}}\ket{\psi}=\ket{\psi}\
\mbox{for any closed path}\ (j_0,\ldots,j_{p-1},j_0),
\medskip\\
\tilde{C}_{j_0,\ldots,j_{p-1}} \,=\,
i^p \tilde{A}_{j_0,j_1}\tilde{A}_{j_1,j_2}\cdots
\tilde{A}_{j_{p-2},j_{p-1}}\tilde{A}_{j_{p-1},j_0}.
\end{array}
\end{equation}
The stabilizer operators $\tilde{C}_{j_0,\ldots,j_{p-1}}$ are Hermitian and
can be represented in the form $\pm\prod_{(j,k)}\sigma^{\alpha_{jk}}_{jk}$,
where $(j,k)$ runs over a set of different qubits. The set of stabilizer
operators is obviously redundant but it is consistent, meaning that (i) they
commute with each other, and (ii) whenever the product of several stabilizer
operators is a constant, this constant is $1$. The number of independent
stabilizer operators equals the number of linearly$\pmod{2}$ independent
cycles which in turn equals $u-m+1$. Hence
\begin{equation}
\dim\calL = 2^{u-(u-m+1)} = 2^{m-1} = \dim\calH_0.
\end{equation}

The operators $\tilde{B}_k$, $\tilde{A}_{jk}$ commute with
$\tilde{C}_{j_0,\ldots,j_{p-1}}$, so they leave the code subspace
invariant. The restrictions of these operators, $\tilde{B}_k|_\calL$ and
$\tilde{A}_{jk}|_\calL$, satisfy the relations analogous to
(\ref{simple_relations})--(\ref{parity_relation}). Thus the correspondence
$B_k\mapsto\tilde{B}_k|_\calL$,\, $A_{jk}\mapsto\tilde{A}_{jk}|_\calL$ extends
to a $*$-algebra homomorphism $\mu:\LL(\calH_0)\to\LL(\calL)$. But
$\dim\calL=\dim\calH_0$, hence $\mu$ is an isomorphism. It can be represented
as $\mu(X)=JXJ^\dag$, where $J:\calH_0\to\calL$ is a unitary map which is
unique up to an overall phase.
\smallskip

Now that the main construction is complete, we can give exact rules for
converting the allowed (geometrically local) operations on LFMs into qubit
gates. These rules are almost obvious but still worth putting them into a
rigorous form. For most generality, consider a two-LFM gate application
$U\fapp{j,k}$, where $(j,k)\in E$. Here $U$ is a physical operator acting on 2
LFMs ($=$~4~Majorana fermions $=$~2~qubits), so it can be expressed in terms
of $A=-ic_0c_2=-\sigma^y\qapp{0}\sigma^x\qapp{1}$,\,
$B'=-ic_0c_1=\sigma^z\qapp{0}$ and $B''=-ic_2c_3=\sigma^z\qapp{1}$. Applying
$U$ to LFMs $j$ and $k$ means substituting $A_{jk}$, $B_j$, $B_k$ for $A$,
$B'$, $B''$. Instead of that, we actually do another substitution:
\begin{equation}
\nu_{jk}\,:\quad
A\mapsto\tilde{A}_{jk},\quad B'\mapsto\tilde{B}_j,\quad B''\mapsto\tilde{B}_k.
\end{equation}
It extends to a $*$-algebra homomorphism $\calG\to\LL(\qb^{\otimes S_{jk}})$,
where $\calG$ is the algebra generated by $A$, $B'$ and $B''$ ($=$~the
algebra of parity-preserving operators on two qubits), and $S_{jk}$ is the set
of qubits consisting of all edges incident to $j$ and $k$. (Note that we do
not have to restrict $\tilde{A}_{jk}$, $\tilde{B}_j$, $\tilde{B}_k$ to the
subspace $\calL$ because the cycle relation (\ref{cycle_relation}) is
irrelevant in this context). So, $U\fapp{j,k}$ is simulated by $\nu_{jk}(U)$.

It turns out that this simulation is pretty efficient even if $d$ ($=$~the
largest degree of a vertex in the graph) is not a constant. W.\,l.\,o.\,g.\
$m>2$, so $S_{jk}$ contains at least one edge besides $(j,k)$, say, $(j,l)$.
Then $\tilde{B}_j$, $\tilde{B}_k$ and $\tilde{A}_{jk}$ have no nontrivial
relations (like $\tilde{B}_j\tilde{B}_k=1\,$). More exactly, the map
$\nu_{jk}$ is injective, so $\tilde{B}_j$, $\tilde{B}_k$ and $\tilde{A}_{jk}$
satisfy exactly the same relations as $\sigma^z\qapp{0}$, $\sigma^z\qapp{1}$
and $-\sigma^y\qapp{0}\sigma^x\qapp{1}$. It follows that there is a
``symplectic transformation''~\cite{Kit2} of the form $X\mapsto WXW^\dag$
which takes $\tilde{B}_j$, $\tilde{B}_k$ and $\tilde{A}_{jk}$ to
$\sigma^z_{jl}$, $\sigma^z_{jk}$ and $-\sigma^y_{jl}\sigma^x_{jk}$. The
unitary operator $W$ acts on the qubits from $S_{jk}$ and can be represented
as a product of elementary ``symplectic gates'': $H$, $\Lambda(\sigma^x)$ and
$\Lambda(i)$. It is easy to show that $O(d)$ applications of these gates are
sufficient. Hence executing the operator
\begin{equation}
\nu_{jk}(U) \,=\ W^\dag\ U\qapp{(j,l),(j,k)}\ W
\end{equation}
takes $O(d)$ one-qubit and two-qubit gate applications.
\smallskip

In the above analysis, we did not take into account the number of operations
required to prepare an initial state $\ket{\psi}\in\calL$, which is necessary
to begin simulation. For definiteness, we will consider the unique quantum
state $\ket{\xi}\in\calL$ which satisfies additional constraints:
\begin{equation}
\tilde{B}_k\ket{\xi}=\ket{\xi}\ \mbox{for each}\ k.
\end{equation}
Note that $B_k=-ic_{2k}c_{2k+1}=1-2a_k^\dag a_k$, so $\ket{\xi}$ represents
the fermionic state $\ket{0,\ldots,0}$. By a general argument, the qubit state
$\ket{0}\otimes\cdots\otimes\ket{0}$ can be transformed into $\ket{\xi}$ by
applying $O(u^2)$ symplectic gates.

However, if we continue the speculations about fermions being possibly
mimicked by bosons, the vacuum state of the bosonic system must absorb new
degrees of freedom as the Universe expands. Certainly, this process should be
reversible, i.\,e.\ it should also work when the space shrinks. In our model,
shrinking the space corresponds to contracting some edges. More specifically,
contracting an edge $(j,k)$ means removing it while identifying the vertices
$j$ and $k$. If both $j$ and $k$ are connected to the same vertex $l$, a
double edge between $j\equiv k$ and $l$ appears; it must be then reduced to a
single one. We should be able to update our ``vacuum state'' $\ket{\xi}$
through these transformations. (The qubit being removed should come out in the
state $\ket{0}$). One can show that a single contraction or reduction step
requires $O(d)$ symplectic gates. This does not involve any nonlocality and
can be done simultaneously in different places, which is quite consistent with
the idea of adiabatic vacuum transformation in the expanding Universe.

\section{Quantum codes by Majorana fermions}

Some symplectic codes {\em on qubits} can be conveniently described in terms
of Majorana fermions. We will show how it works for the Shor
code~\cite{Shor_code}. Whether this approach can help to find new codes still
remains to be seen.

By inverting the formula~(\ref{MFdef}), we can represent the operators
$a_k,a_k^\dag$ in terms of the Majorana operators:
$a_k=\frac{1}{2}(c_{2k}+ic_{2k+1})$,\,
$a_k^\dag=\frac{1}{2}(c_{2k}-ic_{2k+1})$. However, one can also introduce
another set of annihilation and creation operators which will satisfy the same
commutation relations:
\begin{equation}
b_k=\frac{1}{2} \Bigl(c_{\tau(2k)}+ic_{\tau(2k+1)}\Bigr), \qquad
b_k^\dag= \frac{1}{2}\Bigl(c_{\tau(2k)}-ic_{\tau(2k+1)}\Bigr),
\end{equation}
where $\tau:\{0,\ldots,2m\!-\!1\}\to\{0,\ldots,2m\!-\!1\}$ is arbitrary
permutation on $2m$ elements. One can define a quantum code by fixing the
occupation numbers of some of the new LFMs, e.\,g.\ by requiring that each
codevector $\ket{\psi}$ satisfies $b_k^\dag b_k\ket{\psi}=0$ for $k=1\ldots
m-1$. In other words, the set of stabilizer operators is
\begin{equation}
X_k=-i\,c_{\tau(2k)}\,c_{\tau(2k+1)},\quad\ k=1,\ldots m-1.
\end{equation}
The number of encoded qubits is $m-(m-1)=1$. The logical operators for this
code are generated by $c_{\tau(0)}$ (the encoded $\sigma^x$) and $c_{\tau(1)}$
(the encoded $\sigma^y$).

Thus each permutation $\tau:\{0,\ldots,2m\!-\!1\}\to\{0,\ldots,2m\!-\!1\}$
defines a quantum code which encodes $1$ qubit into $m$ qubits. It turns out
that these codes can have arbitrary large code distances. We are to define a
family of such codes which can be considered as slightly modified Shor codes.
Let $l\ge 2$ be an integer (the index of a code in the family), $m=l^2$.
There will be two types of stabilizer operators:
\begin{equation}
\begin{array}{rcl@{\qquad}l}
Z_k &=& -i\,c_{2kl+1}\,c_{2(k+2)l-2}\,, & k=0\ldots l-2,\medskip\\
Y_{k,j} &=& -i\,c_{2kl+2j+3}\,c_{2kl+2j}\,, &
k=0,\ldots l-1,\quad j=0,\ldots l-2.
\end{array}
\end{equation}
They can be expressed in terms of
the Pauli operators as follows~:
\begin{equation}
Z_k \,=\, \sigma^x\qapp{kl}\, \sigma^x\qapp{(k\!+\!2)l\!-\!1}
\prod_{s=kl+1}^{(k+2)l-2}\sigma^z\qapp{s}\,,\qquad
Y_{k,j} \,=\, \sigma^y\qapp{kl\!+\!j}\, \sigma^y\qapp{kl\!+\!j\!+\!1}
\end{equation}

For example, let us consider the $l=3$ code. Its 9 qubits can be arranged into
a $3\times3$ array as follows
\[
\begin{array}{ccc} 0,&1,&2,\\ 3,&4,&5,\\ 6,&7,&8. \end{array}
\]
Then the stabilizer operators of the first type become:
\[
Z_0 \,= \left( \begin{array}{ccc}
\sigma^x, & \sigma^z, & \sigma^z, \\
\sigma^z, & \sigma^z, & \sigma^x, \\
I, & I, & I\phantom{,}
\end{array} \right) , \quad
Z_1 \,=\left( \begin{array}{ccc}
I, & I,& I, \\
\sigma^x, & \sigma^z, & \sigma^z, \\
\sigma^z, & \sigma^z, & \sigma^x\phantom{,}
\end{array} \right).
\]
(These are not matrices. We just mean that, for example,
$Z_0=\sigma^x\otimes\sigma^z\otimes\sigma^z\otimes
\sigma^z\otimes\sigma^z\otimes\sigma^x\otimes
I\otimes I\otimes I$). The stabilizer operators of the second type are
\[
\begin{array}{c}
Y_{0,0} = \left( \begin{array}{ccc}
\sigma^y,& \sigma^y,&I,\\ I,&I,&I, \\ I,&I,&I\phantom{,}
\end{array} \right), \quad
Y_{0,1} = \left( \begin{array}{ccc}
I,& \sigma^y,& \sigma^y,\\  I,&I,&I, \\ I,&I,&I\phantom{,}
\end{array} \right), \quad
Y_{1,0} = \left( \begin{array}{ccc}
I,&I,&I, \\ \sigma^y,&\sigma^y,& I,\\ I,&I,&I\phantom{,}
\end{array} \right), \bigskip\\
Y_{1,1} =\left( \begin{array}{ccc}
I,&I,&I, \\  I,& \sigma^y,&\sigma^y,\\ I,&I,&I\phantom{,}
\end{array} \right), \quad
Y_{2,0}=\left( \begin{array}{ccc}
I,&I,&I, \\  I,&I,&I, \\  \sigma^y,&\sigma^y,&I\phantom{,}
\end{array} \right), \quad
Y_{2,1}=\left( \begin{array}{ccc}
I,&I,&I, \\  I,&I,&I, \\ I,&\sigma^y,&\sigma^y\phantom{,}
\end{array} \right). \\
\end{array}
\]
If one performs the cyclic permutation
$\sigma^x\mapsto\sigma^y\mapsto\sigma^z\mapsto\sigma^x$, the operators
$Y_{k,j}$ turn into certain stabilizer operators of the Shor code. The
operators $Z_{k}$ will not become exactly the same as in the Shor code. Each
of them will contain two $\sigma^y$ instead of $\sigma^x$. However, the code
distance of this code is the same, namely, $3$.

For arbitrary $l$, the code distance is $d_l=l$. (The proof is essentially the
same as for the Shor code).


\begin{thebibliography}{99}
 
\bibitem{Feynman} R.\,P.\,Feynman, 
``Quantum mechanical computers'',
{\it Optics News} {\bf 11}, 11 (1985).

\bibitem{Deutch} D.\,Deutsch,
``Quantum computational networks'',
{\it Proc.\ Roy.\ Soc.\ Lond.}, Ser.~A, {\bf 425}, 73--90 (1989).
 
\bibitem{Lloyd} S.\,Lloyd,
``Unconventional Quantum Computing Devices,''
in {\it Unconventional Models of Computation,}
C.S. Calude, J. Casti, M.J. Dinneen, eds., Springer, Singapore, 1998.

\bibitem{Pfaffian} G.~Moore, N.~Read,
``Nonabelians in the fractional quantum Hall effect'',
{\it Nucl.\ Phys} {\bf B 360}, 362--396 (1991);\\
C.\,Nayak, F.\,Wilczek,
``$2n$-quasihole states realize $2^{n-1}$-dimensional spinor braiding
statistics in paired quantum Hall states'',
{\it Nucl.\ Phys} {\bf B 479}, 529--553 (1996).

\bibitem{Kit1} A.\,Yu.\,Kitaev,
``Fault-tolerant quantum computation by anyons'',
ArXiv: \mbox{\tt quant-ph/9707021}.

\bibitem{AL} D.\,S.\,Abrams and S.\,Lloyd,
``Simulation of many-body fermi systems on a universal quantum computer'',
{\it Phys.\ Rev.\ Lett.}\ {bf 79}, 2586--2589 (1997)
(ArXiv: \mbox{\tt quant-ph/9703054}).

\bibitem{TQFT} M.\,H.\,Freedman, A.\,Kitaev, Z.\,Wang,
``Simulation of topological field theories by quantum computers'',
ArXiv: \mbox{\tt quant-ph/0001071}.

\bibitem{universality} E.\,Knill, R.\,Laflamme, W.\,H.\,Zurek,
``Resilient quantum computation: error models and thresholds'',
{\it Proc.\ Roy.\, Soc.\ Lond.}, Ser.~A, {\bf 454},
365--384 (1998) (ArXiv: \mbox{\tt quant-ph/9702058};\\
P.\,O.\,Boykin, T.\,Mor, M.\,Pulver, V.\,Roychowdhury, F.\,Vatan,
``On universal and fault-tolerant quantum computing'',
ArXiv: \mbox{\tt quant-ph/9906054}.

\bibitem{q-wire} A.\,Kitaev,
``Unpaired Majorana fermions and protected quantum memory'',
In preparation.
 
\bibitem{Ising} A.\,Kitaev, S.\,Bravyi,
In preparation.

\bibitem{AKN} D.\,Aharonov, A.\,Kitaev, N.\,Nisan,
``Quantum Circuits with Mixed States'',
{\it Proc.\ 30th Annual ACM Symposium on Theory of Computation (STOC)},
20--30 (1997) (ArXiv: \mbox{\tt quant-ph/9806029}).

\bibitem{sympcodes} A.\,R.\,Calderbank, E.\,M.\,Rains,
P.\,W.\,Shor, N.\,J.\,A.\,Sloane, 
``Quantum error correction and orthogonal geometry'',
{\it Phys.\ Rev.\ Lett.}\ {\bf 78}, 405-408 (1997)
(ArXiv: \mbox{\tt quant-ph/9605005}).

\bibitem{Kit2} A.\,Yu.\,Kitaev,
``Quantum computations: algorithms and error correction'',
{\it Russian Math.\ Surveys} {\bf 52:}6, 1191--1249 (1997)
(original Russian version: Uspekhi Mat.\ Nauk {\bf 52:}6, 53--112).

\bibitem{Shor_code} P.\,W.\,Shor,
``Schemes for reducing decoherence in quantum computer memory'',
{\it Phys.\ Rev.}\ {\bf A 52}, 2493--2496 (1995).


\end{thebibliography}
\end{document}